\documentclass[twocolumn,superscriptaddress,amsmath,amssymb,prl,linenumbers]{revtex4}

\usepackage[dvips]{graphicx}
\usepackage{dcolumn}
\usepackage{bm}
\usepackage{color}

\newcommand {\ryx}{\rho_{yx}}
\newcommand {\rxx}{\rho_{xx}}

\newcommand {\muz}{\mu_{\mathrm{0}}}
\newcommand {\R}{R_{\mathrm{0}}}

\newcommand {\SH}{R_{\mathrm{A}}}
\newcommand {\HC}{H_{\mathrm{c}}}

\newcommand {\rn}{\rho_{yx}^{\mathrm{N}}}
\newcommand {\ra}{\rho_{yx}^{\mathrm{A}}}
\newcommand {\rt}{\rho_{yx}^{\mathrm{T}}}

\begin{document}
\title{Topological magnetic phase in the candidate Weyl semimetal CeAlGe}

\author{Pascal Puphal}
\email{pascal.puphal@psi.ch}
\affiliation{Laboratory for Multiscale Materials Experiments (LMX), Paul Scherrer Institute (PSI), CH-5232 Villigen, Switzerland}
\author{Vladimir Pomjakushin}
\affiliation{Laboratory for Neutron Scattering and Imaging (LNS), Paul Scherrer Institute (PSI), CH-5232 Villigen, Switzerland}
\author{Naoya Kanazawa}
\affiliation{Department of Applied Physics, University of Tokyo, Tokyo 113-8656, Japan}
\author{Victor Ukleev}
\affiliation{Laboratory for Neutron Scattering and Imaging (LNS), Paul Scherrer Institute (PSI), CH-5232 Villigen, Switzerland}
\author{Dariusz J. Gawryluk}\thanks{On leave from Institute of Physics, Polish Academy of Sciences, Aleja Lotnikow 32/46, PL-02-668 Warsaw, Poland}
\affiliation{Laboratory for Multiscale Materials Experiments (LMX), Paul Scherrer Institute (PSI), CH-5232 Villigen, Switzerland}
\author{Junzhang Ma}
\author{Muntaser Naamneh}
\author{Nicholas C. Plumb}
\affiliation{Swiss Light Source (SLS), Paul Scherrer Institute (PSI), CH-5232 Villigen PSI, Switzerland}
\author{Lukas Keller}
\affiliation{Laboratory for Neutron Scattering and Imaging (LNS), Paul Scherrer Institute (PSI), CH-5232 Villigen, Switzerland}
\author{Robert Cubitt}
\affiliation{Institut Laue-Langevin (ILL), 71 avenue des Martyrs, CS 20156, 38042 Grenoble cedex 9, France}
\author{Ekaterina Pomjakushina}
\affiliation{Laboratory for Multiscale Materials Experiments (LMX), Paul Scherrer Institute (PSI), CH-5232 Villigen, Switzerland}
\author{Jonathan S. White}
\email{jonathan.white@psi.ch}
\affiliation{Laboratory for Neutron Scattering and Imaging (LNS), Paul Scherrer Institute (PSI), CH-5232 Villigen, Switzerland}
\begin{abstract}
We report the discovery of topological magnetism in the candidate magnetic Weyl semimetal CeAlGe. Using neutron scattering we find this system to host several incommensurate, square-coordinated multi-$\vec{k}$ magnetic phases below $T_{\rm{N}}$. The topological properties of a phase stable at intermediate magnetic fields parallel to the $c$-axis are suggested by observation of a topological Hall effect. Our findings highlight CeAlGe as an exceptional system for exploiting the interplay between the nontrivial topologies of the magnetization in real space and Weyl nodes in momentum space.
\end{abstract}
\maketitle
%

The recent experimental discoveries of topological Weyl semimetals with discrete band touching points near a small Fermi surface (Weyl nodes) \citep{Xu(2015),Lv(2015),Xu2(2015),Hasan(2015),Hasan(2017)}, augments the viral interest in topological electronic phases of matter~\citep{Bin2017,Arm2018}. In this context, \emph{magnetic} Weyl semimetals are special since they promise topological states that are easily tunable using low external magnetic fields. In ferromagnetic (FM) Weyl semimetals the magnetization ($M$) can modify the pattern of Weyl nodes into a symmetry-breaking configuration that generates a large anomalous Hall effect (AHE)~\citep{Liu2018,Chang(2018),Kim2018}. In magnetic insulators, Weyl magnon states can emerge {\color{red}\citep{Moo16,Li16,Li17,Yao2018,Bao2018}} that generate anomalous heat transport. In each case, the topological origin of anomalous transport augments an anticipation that Weyl systems may find use for topological electronics~\citep{Tok(2017)}.\par

Topologically nontrivial magnetization textures in real-space are also known to generate experimental signatures in transport such as the so-called topological Hall effect (THE). A THE can emerge if the underlying noncoplanar magnetization texture is described by a finite topological number $Q=(1/4\pi)\int\vec{n}\cdot\left(\partial\vec{n}/\partial x\times \partial\vec{n}/\partial y\right)dxdy$ in the continuous limit. Here unit vector $\vec{n}$ describes the direction of the local magnetization. Several magnets are known to host topological skyrmion ($Q$=-1)~\citep{Muh2009,Sek12,Kez15,Tok15}, biskyrmion ($Q$=-2)~\citep{Yu14,Wang2016,Li2019}, antiskyrmion ($Q$=+1)~\citep{Nay17}, and meron-antimeron ($Q=\pm$1/2)~\citep{Yu18} magnetic textures. For those with $|Q|$ described by integer $\mathbb{Z}\geq1$ on a macroscopic scale, the THE is indeed observed in favourable cases~\citep{Neu09,Kan11,Sch12,Wang2016,Kan17}.\par

Natural questions concern if novel functions can emerge due to a coexistence of topological magnetism with Weyl nodes. For example, the large spin-orbit coupling in Weyl semimetals can promote spin transfer torques, in turn enabling an energy efficient manipulation of topological magnetism by electric currents~\citep{Tak16,Kur2019}. Up to now the account of magnetic Weyl semimetals has focussed mainly on systems with commensurate collinear order~\citep{Liu2018,Chang(2018),Kim2018}; expansion into the realm of semimetals with incommensurate (IC) noncoplanar magnetism is of considerable interest, yet has remained hitherto unexplored.\par

\begin{figure}
\includegraphics[width=1\columnwidth]{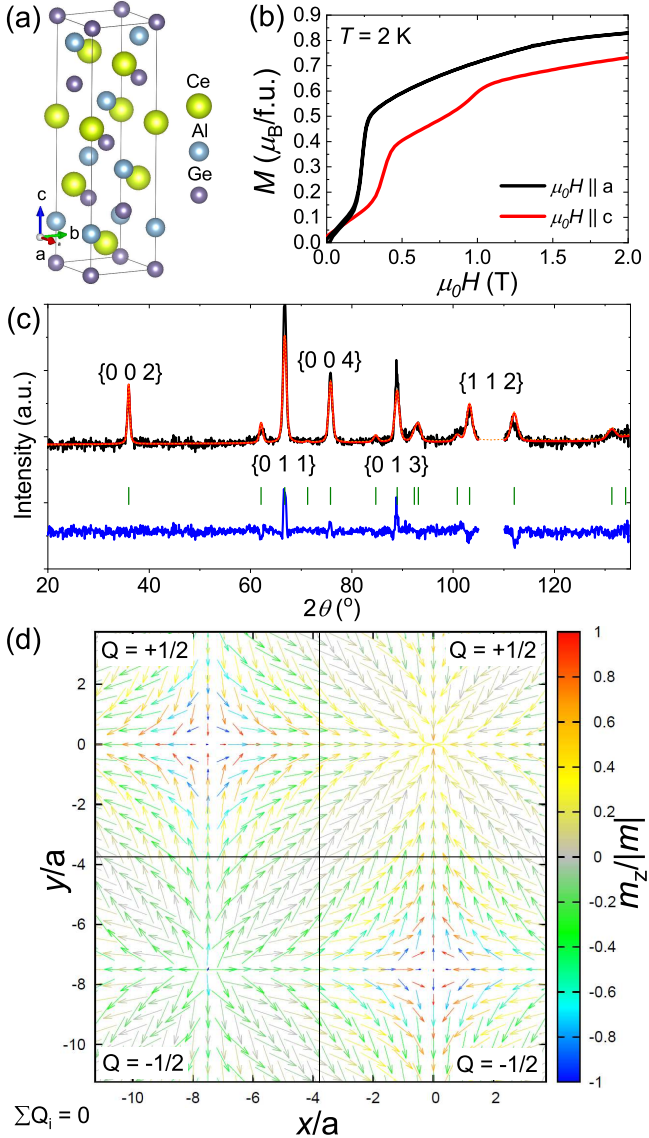}
\caption{(a) $I$4$_{1}$md crystal structure of CeAlGe (b) $M$ vs. $\mu_{0}H$ at 2 K after ZFC (c) Magnetic NPD difference profile taken between $T=1.7$\,K and 10\,K, at $\mu_{0}H$=0 (black). Red shows the Rietveld refinement of the data using the multi-$\vec{k}$ model (see text for details). Blue shows the difference between the data and refinement. Green markers indicate scattering angles for possible IC magnetic peaks. (d) View along the $z$-($c$-)axis of the normalized (i.e. $\vec{n}=\vec{M}/|\vec{M}|$, where $\vec{M}$ is the local Ce moment) multi-$\vec{k}$ magnetic structure. $x$- and $y$-axes are in units of $a$-axis lattice parameter. The length of arrows denote the size of in-plane $\vec{n}$. The colorbar shows the value of $\vec{n}_{z}$. Black quadrants highlight regions with $Q=$+1/2 or -1/2.}
\label{Fig_1}
\end{figure}

Recently the polar tetragonal magnet CeAlGe ($I$4$_{1}$md spacegroup, $C_{4v}$ symmetry [Figure~\ref{Fig_1}(a)]) was predicted to be an easy-plane FM type-II Weyl semimetal \citep{Chang(2018)}. In contrast to FM order however, bulk measurements show magnetic Ce atoms instead undergo an antiferromagnetic (AFM)-like transition below $T_{\rm{N}}$ $\sim$4.4\,K, with an estimated in-plane AFM interaction of $\sim$-42\,K, and an out-of-plane FM interaction of $\sim$10\,K~\citep{Puphal(2019)}. From the magnetic field- ($\mu_{0}H$) dependent bulk magnetization $M$ the easy-plane anisotropy is confirmed [Figure~\ref{Fig_1}(b)], but unexpected low field anomalies imply richer magnetism than simple FM order~\citep{Puphal(2019)}. In addition, $C_{nv}$ polar magnets are known in theory~\citep{Bog89} to allow Dzyaloshinskii-Moriya interactions (DMIs) that stabilize N\'{e}el-type skyrmion phases \citep{Kez15,Kur2017,Bor17}. Thus the candidate Weyl semimetal CeAlGe is also a promising host of topological magnetism.\par

In this Letter, we apply neutron scattering and electrical transport measurements to reveal CeAlGe as a host of IC multi-$\vec{k}$ magnetic phases, including a topological one at intermediate $\mu_{0}H$$\parallel$$c$ containing pairs of $Q$=+1/2 antimerons. We discuss the implication for the existence of such IC magnetic phases in Weyl semimetals suggesting they may act as a novel platform for new functions.\par

Neutron powder diffraction (NPD) was done on annealed polycrystalline CeAlGe samples using the DMC and HRPT instruments at PSI. Full experimental details are given in Ref.~\onlinecite{Sup}. At HRPT, we tracked the $T$-dependent crystal structure during zero field-cooling (ZFC), and find CeAlGe to retain its room $T$ polar $I4_{1}$md crystal structure down to 1.5\,K~\citep{Sup}. Figure~\ref{Fig_1}(c) shows the difference between NPD data acquired at DMC below (1.5~K) and above (10~K) $T_{\rm{N}}$. A Le Bail refinement of the data~\citep{Sup} shows all magnetic peaks can be indexed by $\vec{G}+\vec{k}$. Here $\vec{G}$ is a reciprocal lattice vector, and $\vec{k}=(a,0,0)$ is an IC magnetic propagation vector with $a$=0.066(1) in reciprocal lattice units (r.l.u.). With support from single crystal data presented later, CeAlGe is found to have a purely IC magnetic ground state.\par

To determine the IC magnetic structure in CeAlGe we used the Fullprof suite \citep{Rodriguez-Carvajal1993} to perform a detailed Rietveld refinement of the NPD difference pattern against several symmetry-allowed models~\citep{Sup}. The symmetry analysis shows that the magnetic structure model of highest symmetry is based on the full propagation star of $\vec{k}$ which consists of four arms $\vec{k_{1}}$=$\pm(a,0,0)$ and $\vec{k_{2}}$=$\pm(0,a,0)$. This corresponds to a multi-$\vec{k}$ model described by the maximal symmetry superspace group $I4_{1}$md1'(a00)000s(0a0)0s0s \cite{Campbell2006,Aroyo2006}, and has just four refinable magnetic mode amplitude parameters~\citep{Sup}. Figure~\ref{Fig_1}(c) shows the refinement ($\chi^{2}$=2.48) of the NPD difference pattern using the multi-$\vec{k}$ model~\citep{Sup}, with the reconstructed magnetic order presented in Figure~4 of Ref.~\onlinecite{Sup}. For clarity, in Figure~\ref{Fig_1}(d) we show the reconstructed magnetic order with normalized moments, $\vec{n}=\vec{M}/|\vec{M}|$. The magnetic structure modulates only within the tetragonal plane and, as discussed later, is characterized by quadrants for which $Q$ is modulo half-integer, with $Q$=0 on a macroscopic scale.\par

In the Supplement~\citep{Sup}, we also consider data refinements against models generated by a standard representation analysis without magnetic group symmetry arguments. In this approach just a single arm of $\vec{k}$ is considered, necessarily leading to models of lower symmetry than the multi-$\vec{k}$ model. Here the 4$a$ Ce atom splits into two orbits Ce$_{1}$ and Ce$_{2}$, with their moments being symmetry-independent. The best refinement is achieved by a six parameter model that describes $a$-$c$ cycloids on each Ce orbit, with $\chi^{2}=2.48$ the same as for the multi-$\vec{k}$ model. Thus while strictly unable to rule out that a multidomain, single-$\vec{k}$ cycloidal order forms the ground state in zero field, we prioritize the multi-$\vec{k}$ model due to both its higher symmetry and fewer refinement parameters.\par


\begin{figure}
\includegraphics[width=1\columnwidth]{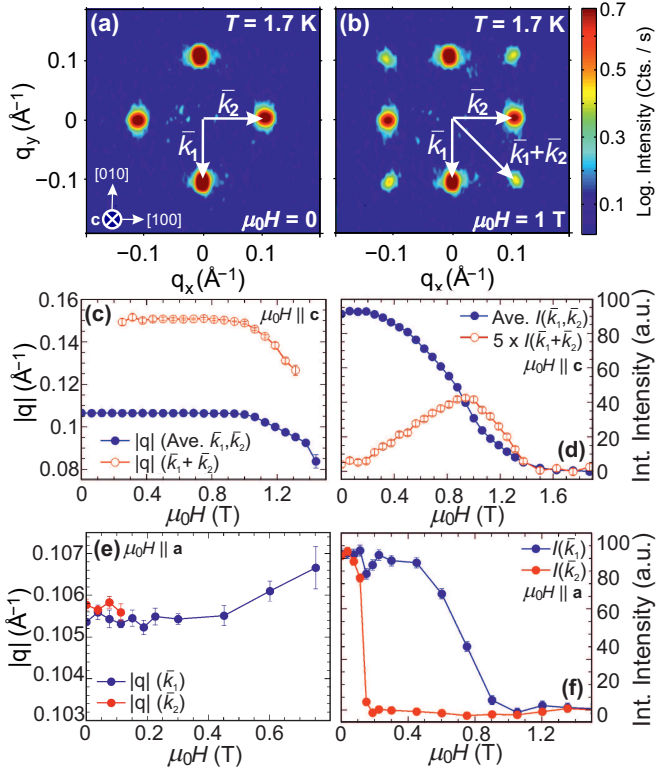}
\caption{SANS patterns from single crystal CeAlGe at 1.7~K and (a) $\mu_{0}H$=0, (b) $\mu_{0}H$$\parallel$$c$=1~T. Note the logarithmic intensity scale. The $\mu_{0}H$$\parallel$$c$-dependence of (c) IC peak wavevectors $|q|$ and (d) peak integrated intensities at 1.7~K. Each datapoint is an average taken over equivalent wavevectors. (e)-(f) show similar data as for (c)-(d) but for $\mu_{0}H$$\parallel$$a$ at 1.9~K.}
\label{Fig_2}
\end{figure}


Next we consider $\mu_{0}H$-dependent studies on stoichiometric floating zone grown single crystal CeAlGe~\citep{Puphal(2019),Mis2}. Figure~\ref{Fig_2} shows data obtained by small-angle neutron scattering (SANS) at SANS-I, PSI, and D33 at the ILL~\citep{Sup}, which probes directly the IC magnetic modulations near $q$=0~\citep{Mis1}. After ZFC to 1.7~K, the SANS pattern shown in Figure~\ref{Fig_2}(a) displays the four IC Bragg peaks expected from all arms of $\vec{k}$. Within uncertainty, all peaks display absolute wavevectors of $|q|$=0.107(1)~\AA$^{-1}$, corresponding to $\vec{k}=(a,0,0)$, where $a$=0.071(1) r.l.u., this being slightly larger than for the powder. As exemplified by data obtained at 1~T shown in Figure~\ref{Fig_2}(b), with increasing $\mu_{0}H$$\parallel$$c$ SANS intensity emerges at second-order scattering vectors termed $\vec{k_{1}}+\vec{k_{2}}$. The observation of second-order peaks generally implies the field to induce anharmonicity in the magnetic order, and is a hallmark of topological multi-$\vec{k}$ phases~\citep{Ada10,Oka17}.

Figures~\ref{Fig_2}(c) and (d) respectively show the $\mu_{0}H$$\parallel$$c$-dependences of the absolute sizes of the first- and second-order peak wavevectors, and the associated peak integrated intensities. The very different $\mu_{0}H$-dependences of the first-and second-order peak intensities, and the absence of second-order peak intensity at $\mu_{0}H$=0 when the first-order peak intensity is largest, together provide convincing evidence that the second-order signal is intrinsic, and not due to multiple scattering. Importantly, both the first-and second-order peak intensities vary only smoothly with $\mu_{0}H$$\parallel$$c$, showing no discontinuous changes. This suggests that the inherent symmetry of the modes describing the IC texture is the same at finite $\mu_{0}H$$\parallel$$c$ as determined for $\mu_{0}H$=0.\par




\begin{figure}
\includegraphics[width=1\columnwidth]{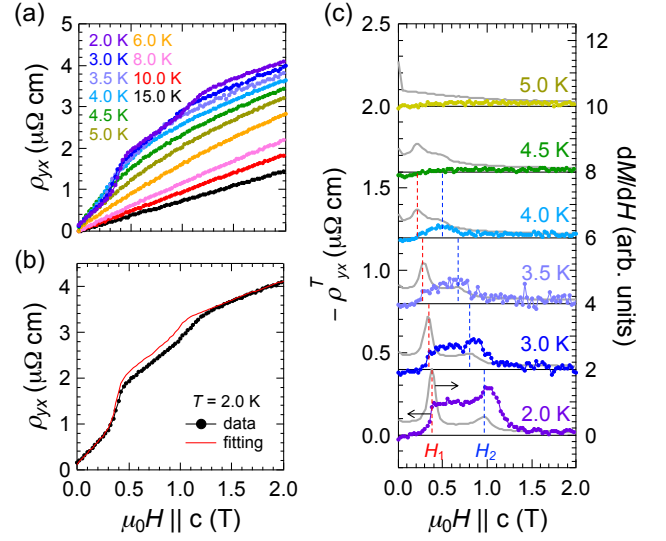}
\caption{(a) $\mu_{0}H$$\parallel$$c$-dependence of the Hall resistivity $\ryx$ at various $T$s. (b) Example of the appearance of the topological Hall resistivity $\rt$ as a deviation from the fitted conventional Hall contributions $\ryx=\muz\R H +\SH \rxx^2 M$ at $T=2$ K. (c) Comparison between the estimated $\rt$ and the field-derivative of $M$. A stepwise $\rt$ appears for $H_1<\mu_{0}H<H_2$, suggesting the formation of topological magnetic order.}
\label{Fig_3}
\end{figure}

For $\mu_{0}H$$\parallel$$a$, Figures~\ref{Fig_2}(e)-(f) show that just a single-$\vec{k}$ IC order described by $\vec{k_{1}}$ prevails above 0.18~T - see also Ref.~\onlinecite{Sup}. The perpendicular alignment of $\vec{k_{1}}$ to $\mu_{0}H$ is consistent with cycloidal order~\citep{Kur2017,Whi18}, suggesting the DMI to be crucial in determining the magnetic textures in this $C_{4v}$ system. A complete suppression of IC order at lower $\mu_{0}H$$\parallel$$a$ compared with $\mu_{0}H$$\parallel$$c$ is consistent with the easy-plane anisotropy of CeAlGe \citep{Puphal(2019),Hodovanets(2018)}.\par


Next we turn to Hall effect data for $\mu_{0}H$$\parallel$$c$ [Figure~\ref{Fig_3}(a)]. An additional contribution to the Hall effect, the THE, arises as a hallmark of the formation of noncoplanar spin textures. We attribute the topological Hall resistivity $\rt$ to be a deviation from the conventional Hall signals, which are the $H$-linear normal Hall effect (NHE) and $M$-linear AHE \citep{Kan11}. Namely, we employ the following relation $\ryx=\rn+\ra+\rt=\muz\R H +\SH \rxx^2 M +\rt,$ where $\rn$ and $\ra$ are normal and anomalous Hall resistivity; $\R$ and $\SH \rxx^2$ are their coefficients, respectively. Since $\rt$ is absent when spins are fully polarized under high magnetic fields ($\mu_{0}H>\mu_{0}\HC$), we determine $\R$ and $\SH$ as the intercept and the slope of the curve $\ryx/H$ vs $\rxx^2 M/H$. As shown Figure~\ref{Fig_3}(b), above $\mu_{0}\HC$ $\ryx$ is well reproduced by the fitting curve $\ryx=\muz\R H +\SH \rxx^2 M$. From the difference between the fit and the data, a finite $\rt$ is identified for the intermediate $\mu_{0}H$$\parallel$$c$-region [Figure~\ref{Fig_3}(c)]. This finite $\rt$ provides compelling evidence that the IC multi-$\vec{k}$ magnetism observed by SANS has topological properties described by integer $Q$. In contrast, analysis of further Hall data collected for $\mu_{0}H$$\parallel$$a$ shows that for this field direction $\ryx$ can be described completely in terms of the NHE and AHE~\citep{Sup}. The absence of any THE-like Hall signal suggests none of the various magnetic phases for $\mu_{0}H$$\parallel$$a$ have topological properties described by integer $Q$.\par

\begin{figure}
\includegraphics[width=1\columnwidth]{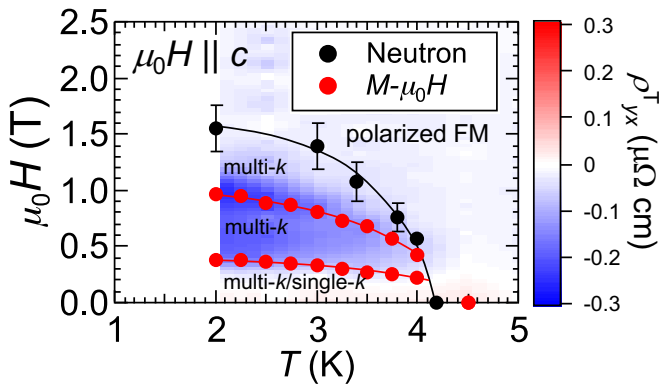}
\caption{CeAlGe magnetic phase diagram for $\mu_{0}H$$\parallel$$c$. Black symbols are determined from SANS data~\citep{Sup}, and red symbols from peaks in d$M$/d$H$ denoted as $H_{1}$ and $H_{2}$ in Figure~\ref{Fig_3}(c). $\rt$ data are included as a colormap. Solid lines are guides for the eye.}
\label{Fig_4}
\end{figure}

Figure~\ref{Fig_4} shows the CeAlGe magnetic phase diagram for $\mu_{0}H$$\parallel$$c$. Phase lines constructed from both neutron data \citep{Sup} and peaks in the field-derivative of bulk $M$-$\mu_{0}H$ isotherms [Figure~\ref{Fig_3}(c)] separate a single-$\vec{k}$/multi-$\vec{k}$ phase for $0<\mu_{0}H<H_{1}$, the intermediate field topological multi-$\vec{k}$ phase for $H_{1}<\mu_{0}H<H_{2}$, and a higher field phase for $H_{2}<\mu_{0}H<H_{c}$. Here $H_{c}$ denotes the saturation field. A finite THE is also observed in the highest field phase [Figure~\ref{Fig_3}(c)], suggesting this phase also to host topological properties that merit further investigation.\par


We now discuss the topological number $Q$ of the lower field IC magnetic phases for $\mu_{0}H$$\parallel$$c$. Usually $Q$ is evaluated in the continuous limit using a coarse grain description of the magnetization texture~\citep{Muh2009,Nag13}. Instead, here we evaluate $Q$ numerically according to complete magnetic structures on the discrete lattice - see Ref.~\onlinecite{Sup} for the details. For the refined multi-$\vec{k}$ model at $\mu_{0}H$=0 [Figure~\ref{Fig_1}(d)], we identify quadrants that are periodic in the tetragonal plane wherein the magnetic order can be characterized by exactly $Q$=+0.50(1) or -0.50(1). If instead the magnetism is a single-$\vec{k}$ cycloid, $Q$=0 within the same quadrants. In either case $Q$=0 on a macroscopic scale, consistent with no low field THE.\par



\begin{figure}
\includegraphics[width=1\columnwidth]{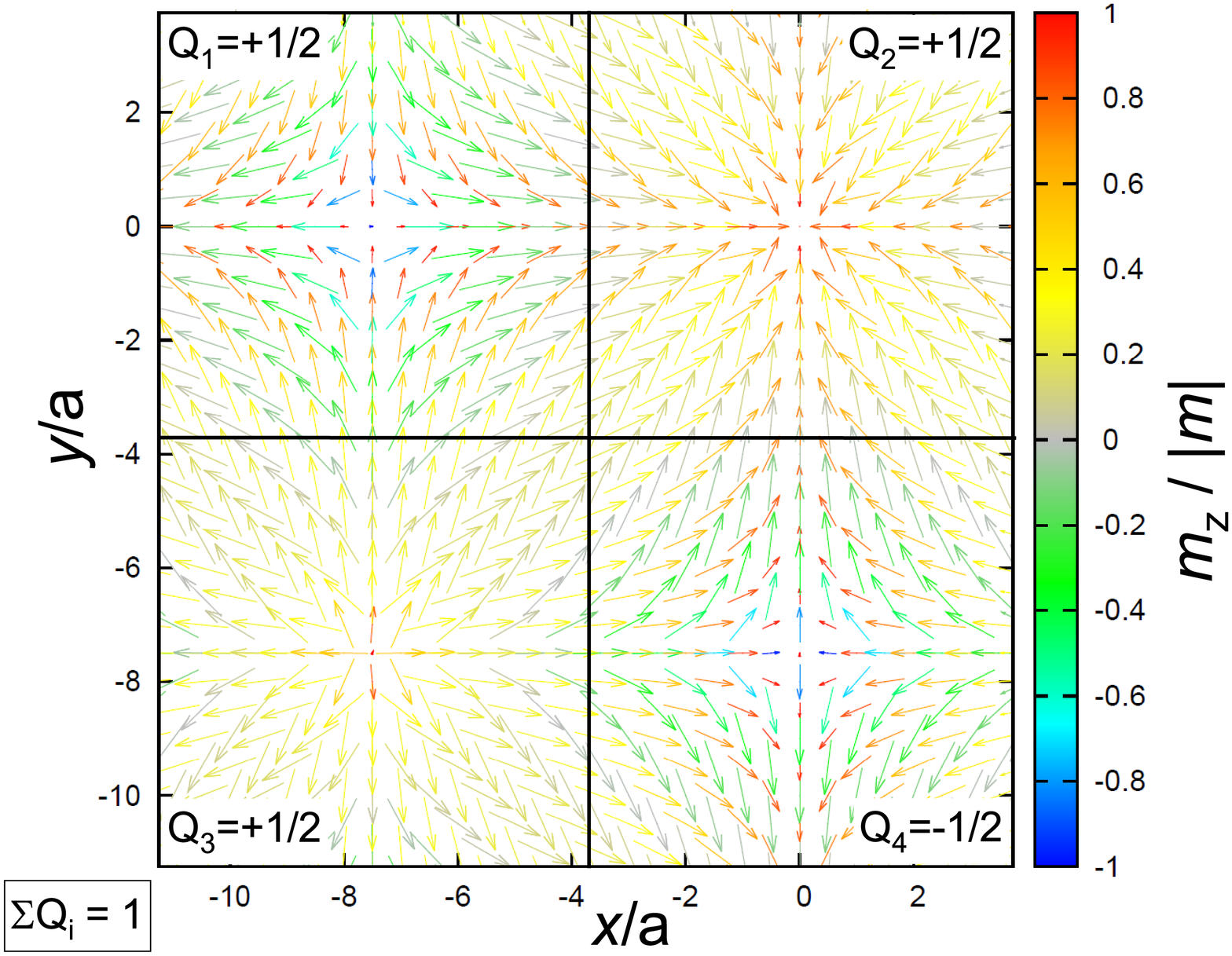}
\caption{View along the $z$-($c$-)axis of the anticipated topological multi-$\vec{k}$ magnetic structure at intermediate $\mu_{0}H$$\parallel$$c$. The magnetic order is presented with normalized Ce moments $\vec{n}=\vec{M}/|\vec{M}|$, similarly as for Figure~\ref{Fig_1}(d).}
\label{Fig_5}
\end{figure}

For the topological phase at intermediate $\mu_{0}H$$\parallel$$c$, a direct determination of the magnetic order requires further studies. Here real-space imaging studies could be helpful. To obtain a candidate magnetic structure model, we use the deduction from SANS that the IC magnetic mode symmetry in finite field is the same as determined for $\mu_{0}H$=0, and add a $\mu_{0}H$-linear induced magnetization $m_{f}$ along $c$ to affect the canting of the refined multi-$\vec{k}$ order in zero field~\citep{Sup}. From this simulation we find that $\sum Q_{i}$ changes sharply between 0 and +1 at $m_{f}$=0.08~$\mu_{\rm{B}}$/Ce, with the anticipated spin texture at intermediate $\mu_{0}H$$\parallel$$c$ visualized in Figure~\ref{Fig_5}. The change to integer $\sum Q_{i}$ which can explain the finite THE, originates from the sign reversal of $Q$ for the quadrant $Q_{3}$ which itself arises due to a reversal of moment components antialigned with $\mu_{0}H$.\par

In the context of topological spin textures, half-integer $Q$ values describe meron ($Q$=-1/2) and antimeron ($Q$=+1/2) textures~\citep{Lin15,Yu18}. Unlike skyrmions where the spin texture wraps a sphere and $|Q|$=$1$, merons and antimerons are described generally by a magnetization aligned up or down along $z$ in a core region, that gradually aligns in-plane towards the edge. Following the classification scheme given in Refs.~\onlinecite{Nag13,Yu18}, the spin textures in the top-right ($Q_{2}$) and bottom-left ($Q_{3}$) quadrants shown in Figure~\ref{Fig_5} are consistent with being core-up antimerons with opposite helicity. The remaining quadrants ($Q_{1}$ and $Q_{4}$) in Figure~\ref{Fig_5} are different to (anti)merons, since they contain moments both aligned \emph{and} antialigned with $z$. For these quadrants the net $Q$ always cancels and they can not contribute to the THE.\par

In finite $\mu_{0}H$, symmetry analysis no longer restricts the \emph{phase} between the IC magnetic modes of the multi-$\vec{k}$ order as it does for $\mu_{0}H$=0, even if though the SANS data imply the inherent mode symmetry remains unchanged. Thus other candidate models for a topological square-coordinated (i.e. double-$\vec{k}$) spin texture can be suggested from theories for easy-plane magnets~\citep{Yi09,Lin15,Oza16}, including meron-antimeron lattices. This contrasts with triple-$\vec{k}$ N\'{e}el-type $|Q|=1$ hexagonal skyrmion lattice phases discovered up to now in bulk polar magnets such as $C_{3v}$ lacunar spinels~\citep{Kez15,Bor17}, and $C_{4v}$ VOSe$_{2}$O$_{5}$~\citep{Kur2017}. In addition, the aforementioned N\'{e}el-type skyrmion magnets are all magnetoelectric insulators with dominant FM interactions. This makes CeAlGe a hitherto unique polar magnet that hosts topological magnetism with a leading AFM interaction, and seemingly without a triple-$\vec{k}$ phase. The description of the distinct magnetic phase diagram of CeAlGe invites a detailed theoretical analysis.\par

Finally we consider the relevance of the complex IC magnetic orders discovered here with respect to the Weyl physics expected in CeAlGe. In magnetic Weyl semimetals, the onset of simple FM order shifts Weyl node separations in $\textbf{k}$-space into a well-defined, time-reversal symmetry breaking configuration~\citep{Chang(2018)}. For spatially modulated IC magnetism, the shifts in node separation can become non-uniform and vary with position in $\textbf{k}$-space. Theory shows that a position-space dependence of Weyl node separations can generate additional gauge fields of Weyl fermions~\citep{Gru16,Pik16}. In turn, these gauge fields may couple to the electronic degrees of freedom with observable consequences at both the microscopic level, by modification of the electronic spectrum, and at the macroscopic level, through generation of a locally enhanced longitudinal conductivity~\citep{Gru16,Pik16}. Our findings thus promote CeAlGe as a potential host of novel functions arising from the direct interplay between Weyl nodes and IC magnetism.


In summary, we identified the polar magnetic Weyl semimetal candidate CeAlGe to host several incommensurate multi-$\vec{k}$ magnetic phases. This includes a topological phase suggested to host antimeron pairs, and which is hitherto unique amongst polar magnets and magnetic Weyl semimetals. While it remains to confirm that Weyl nodes play a key role in the low energy physics of CeAlGe, the discovery of diverse multi-$\vec{k}$ magnetic phases makes this system exceptionally promising for exploring gauge fields of Weyl fermions, and more generally the relationships between magnetically-sensitive, topologically nontrivial features in real- and reciprocal-spaces.\par


V.P. thanks V. Markushin for helpful discussions. J.S.W. acknowledges discussions with A.G.~Grushin and T.~Neupert. Funding from the Swiss National Science Foundation (SNSF) R’Equip Grant Nos. 206021\textunderscore163997 and 206021\textunderscore139082, SNSF projects Nos. 200021\textunderscore159678, and 200021\textunderscore188707, and the SNSF Sinergia network “NanoSkyrmionics” (Grant No. CRSII5\textunderscore171003) is gratefully acknlowedged. This work is based on experiments performed at the Swiss spallation neutron source (SINQ) and the Swiss Light Source (SLS), located at the Paul Scherrer Institute, Villigen, Switzerland, and the Institut Laue-Langevin (ILL), Grenoble, France. This project has received funding from the European Union's Horizon 2020 research and innovation programme under the Marie Skłodowska-Curie grant agreement No. 701647.

\bibliography{cealge}

\begin{thebibliography}{56}
\bibitem[{Mis({\natexlab{b}})}]{Mis1}
\bibinfo{note}{Note that by SANS we only study the IC magnetic scattering,
  since no C positions in reciprocal space were accessible in our experiments.
  Further single crystal data where both IC and C scattering are surveyed
  simultaneously at DMC are given in Ref.~\onlinecite{Sup}, with the results
  entirely consistent with both PND and SANS data.}
\end{thebibliography}
\end{document}